\documentstyle[epsfig]{aipproc}
\begin{document}
\title{Canonical BCS Approximation for the Attractive Hubbard Model}

\author{K. Tanaka and F. Marsiglio}
\address{Department of Physics, University of Alberta,
Edmonton, Alberta, Canada T6G 2J1}

\maketitle

\begin{abstract}
We test the canonical BCS wave functions for fixed number of electrons for 
the attractive Hubbard model. We present results in one dimension for various 
chain lengths, electron densities, and coupling strengths. The ground-state 
energy and energy gap to the first excited state are compared with the exact 
solutions obtained by the Bethe ansatz as well as with the results from the 
conventional grand-canonical BCS approximations.  While the canonical and 
grand-canonical BCS results are both in very good agreement with the exact 
results for the ground state energies, improvements due to conserving the 
electron number in finite systems are manifest in the energy gap.  
As the system size is increased, the canonical results converge to the 
grand-canonical ones.  The ``parity'' effect that arises from the number 
parity (even or odd) of electrons are studied with our canonical scheme.
\end{abstract}

\makeatletter
\global\@specialpagefalse
\def\@oddhead{\hfill Alberta Thy 03-99}
\let\@evenhead\@oddhead
\makeatother


In recent experiments on nanoscale superconducting Al particles
(e.g., Ref.~\cite{black96}), the number of electrons in a particle was fixed
by charging effects and the excitation spectrum has been measured
by tunneling of a single electron.  
They have observed the number ``parity'' effect, the difference 
in the excitation spectra that arises due to the number of electrons
in a particle being even or odd.
For these systems, the validity of the BCS theory that utilizes
the grand canonical ensemble is questionable.
We address this issue by using the attractive Hubbard model, which
serves as the minimal model to describe s-wave superconductivity.
We test the canonical BCS wave function that has a fixed number of 
electrons for this system.
While our formulation can be applied for larger systems up to three dimensions,
we present results for one dimension, for various system sizes, electron
densities and coupling strengths.
Exact solutions are available in one dimension via Bethe Ansatz techniques 
\cite{lieb68}, and the grand canonical BCS solutions have recently been 
evaluated for large system sizes \cite{marsiglio97}.


The attractive Hubbard Hamiltonian is given by
\begin{mathletters}
\begin{eqnarray}
H&=&-\,\sum_{i,\delta \atop \sigma}t_\delta\,(a_{i+\delta,\sigma}^\dagger 
a_{i\sigma} 
+{\rm h.c.}) - |U| \sum_{i} n_{i\uparrow} n_{i\downarrow}
\label{hamilc}\\
&=&\sum_{\bf{k}\sigma} \epsilon_{\bf{k}}\,a_{\bf{k}\sigma}^\dagger
a_{\bf{k}\sigma} - {|U|\over N} \sum_{\bf{k}\bf{k}'\bf{l}}
\,a_{{\bf k}\uparrow}^\dagger
a_{-{\bf k} + {\bf l}\downarrow}^\dagger a_{-{\bf k}' + {\bf l} \downarrow} 
a_{{\bf k}'\uparrow}
\;,\label{hamilk}
\end{eqnarray}
\end{mathletters}
where $a_{i\sigma}^\dagger$ ($a_{i\sigma}$) creates (annihilates) an electron 
with spin $\sigma$ at site $i$ and $n_{i\sigma}$ is the number operator, 
$n_{i\sigma}=a_{i\sigma}^\dagger a_{i\sigma}$
($i$ is the index for the primitive vector ${\bf R}_i$).
The $t_\delta$ is the hopping rate of electrons from one site to a neighbouring
site ${\bf R}_\delta$ away.  In this work, we include the 
nearest neighbours only, and $t\equiv t_\delta$.
The $|U|$ is the coupling strength between electrons on the same site,
and we have explicitly included the fact that the interaction is attractive.
In Eq.(\ref{hamilk}), we have Fourier transformed the Hamiltonian 
with periodic boundary conditions for $N$ sites in each dimension.
The kinetic energy is given by
\begin{equation}
\epsilon_{\bf{k}}=-2\,\sum_{\delta}t_\delta\,{\rm cos}\,{\bf k}\cdot
{\bf R}_\delta\;,
\label{kenergy}
\end{equation}
where ${\bf R}_\delta$ is the coordinate vector connecting sites $i$ to 
$i+\delta$.

We perform variational calculations using the component of the
BCS wave function that has a given number of pairs $\nu$
and thus conserves the number of electrons $N_e=2\nu$.
The BCS wave function is a superposition of pair states with all the possible
numbers of pairs $\{\nu\}$:
\begin{equation}
|{\rm BCS}\rangle_{\rm GC}=\prod_{\bf{k}}\,\bigl(\,u_{\bf k}+v_{\bf k}\,
a_{{\bf k}\uparrow}^\dagger a_{-{\bf k}\downarrow}^\dagger\,\bigr)\,|0\rangle
\equiv \sum_{\nu=0}^{\infty}|\Psi_\nu\rangle\;,
\label{bcswf}
\end{equation}
where $|0\rangle$ denotes the vacuum state, and 
$|\Psi_0\rangle\equiv |0\rangle$.
The $\nu$-pair component $|\Psi_\nu\rangle$ can be obtained by 
rearranging $|{\rm BCS}\rangle_{\rm GC}$ into a power series of the pair 
creation operator $a_{{\bf k}\uparrow}^\dagger a_{-{\bf k}\downarrow}^\dagger$
and can be written as 
\begin{equation}
|\Psi_{2\nu}\rangle={1\over \nu !}\,\prod_{i=1}^{\nu}\,\Bigl(\,
\sum_{{\bf k}_i} g_{{\bf k}_i}\,a_{{\bf k}_i\uparrow}^\dagger 
a_{-{\bf k}_i\downarrow}^\dagger\,\Bigr)\,|0\rangle\;,\label{cbcswfp}
\end{equation}
where ${\bf k}_i\not= {\bf k}_j$ for all $i\not= j$, and we have defined 
$g_{{\bf k}_i}=\Bigl(\prod_{\bf k} u_{\bf k}\Bigr)^{1/\nu}
v_{{\bf k}_i}/u_{{\bf k}_i}$.
An alternative way of formulating the problem is to express the above
wave function as a particle-number projection of $|{\rm BCS}\rangle_{\rm GC}$
\cite{dietrich64,braun98,tanaka99}.
The wave function with an odd number of electrons $N_e=2\nu+1$ is defined by
\begin{equation}
|\Psi_{2\nu+1}\rangle=a_{{\bf q}\sigma}^\dagger\,{1\over \nu !}\,
\prod_{i=1}^{\nu}\,\Bigl(\,\sum_{{\bf k}_i} g_{{\bf k}_i}\,
a_{{\bf k}_i\uparrow}^\dagger a_{-{\bf k}_i\downarrow}^\dagger\,\Bigr)\,
|0\rangle\;,\label{cbcswfop}
\end{equation}
with ${\bf k}_i\not= {\bf k}_j$ for all $i\not= j$.
We construct the energy gap by the definition
\begin{equation}
\Delta_{N_e} = {1\over 2} (E_{N_e - 1} - 2E_{N_e} +  E_{N_e + 1})\;.
\label{cangap}
\end{equation}

\begin{figure}[t!] 
\centerline{\epsfig{file=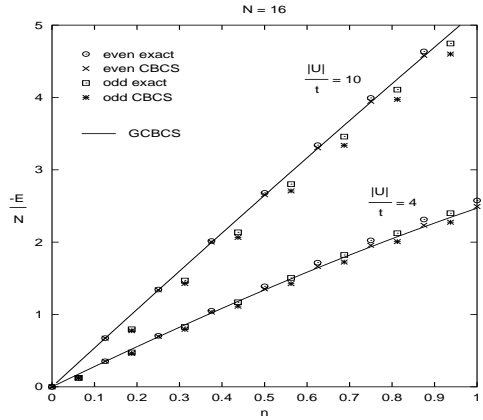,height=2.25in,width=2.5in}}
\vspace{10pt}
\caption{Ground state energy as a function of the electron density for $N=16$.}
\label{fig1}
\end{figure}

In Fig.~\ref{fig1} we show the ground state energy as a function of the
electron density $n=N_e/N$ for $N=16$, for $|U|/t=10$ and 4.
The improvements of the canonical over the grand canonical energies
can be seen, while the former with even $N_e$'s converge to the latter
for smaller $n$ and larger $|U|$, and both results are in very good agreement
with the exact ones for small $n$.
The difference between the energies with even and odd $N_e$'s is apparent
for large $|U|$.

\begin{figure} 
\centerline{\epsfig{file=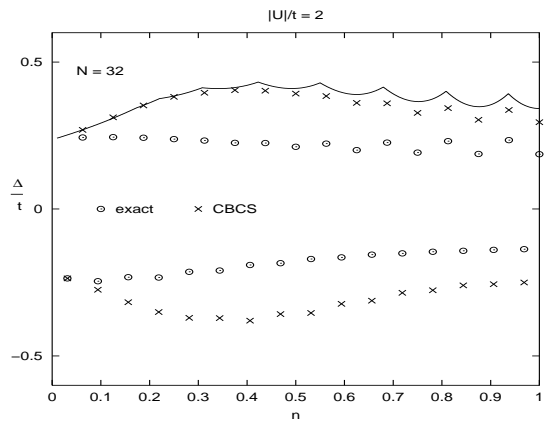,height=2.25in,width=3in}}
\vspace{10pt}
\caption{Energy gap as a function of the electron density for $N=32$ and
$|U|/t=2$.}
\label{fig2}
\end{figure}

The energy gap (scaled by the kinetic energy parameter $t$) is shown
in Fig.~\ref{fig2} as a function of the electron density, 
for $N=32$ and $|U|/t=2$.
The canonical method improves the grand canonical one significantly for
small system size and weak coupling.  As the system size or the coupling
strength is increased, the canonical results converge towards 
the grand canonical ones.  For $|U|/t=2$, $N=32$ is large enough that
the canonical gaps are closer to the grand canonical ones than to the
exact solutions, and the canonical and grand canonical gaps are more or 
less converged for small $n$.
The gap for odd $N_e$ is negative.  Its magnitude is different from 
the even $N_e$ gap for larger density and weak coupling.
Also in this limit, the even gap oscillates as a function of the density.
Both of these effects result from the quantized (unperturbed) energy levels
due to finite size,
and become negligible for strong coupling and for large system size 
(small density).

\begin{figure} 
\centerline{\epsfig{file=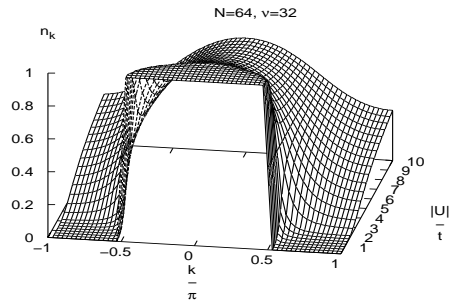,height=2in,width=3in}}
\vspace{10pt}
\caption{Occupation probability for different $k$ states as a function of
the coupling strength for $N=64$ and $\nu=32$.}
\label{fig3}
\end{figure}

Finally, in Fig.~\ref{fig3}, we show the occupation probability of
different $\bf k$ states as a function of the coupling strength.
In the strong coupling limit, all the unperturbed states are more or less
equally occupied (in Fig.~\ref{fig3}, this limit is not reached yet).
As the coupling strength goes to zero, the distribution function for free
electrons is recovered.  Though the figure shown has been produced by the
canonical scheme, the grand canonical approximation also yields the same
probability distribution for such a large system size.

In conclusion, 
the canonical method yields improvements in the energy gap for small
system size and weak coupling.
Further study is being made for comparison with the grand canonical scheme
that conserves the number parity \cite{janko94}.

\subsection*{Acknowledgments}

F. M. acknowledges support from NSERC of Canada, and the Canadian Institute
for Advanced Research.  K. T. acknowledges generous support from the
Avadh Bhatia Fellowship.  We thank Bob Teshima for suggestions to improve
the numerical efficiency for evaluating required integrals.
Calculations were performed on the 42 node SGI
parallel processor at the University of Alberta.

\end{document}